\documentclass[acmtog]{acmart}
\AtBeginDocument{%
  \providecommand\BibTeX{{%
    \normalfont B\kern-0.5em{\scshape i\kern-0.25em b}\kern-0.8em\TeX}}}

\copyrightyear{2023}
\acmYear{2023}
\setcopyright{rightsretained}
\acmConference[EAAMO '23]{Equity and Access in Algorithms, Mechanisms, and Optimization}{October 30-November 1, 2023}{Boston, MA, USA}
\acmBooktitle{Equity and Access in Algorithms, Mechanisms, and Optimization (EAAMO '23), October 30-November 1, 2023, Boston, MA, USA}
\acmDOI{10.1145/3617694.3623226}
\acmISBN{979-8-4007-0381-2/23/11}

\usepackage{multirow}

\usepackage{booktabs}
\usepackage{cleveref}
\usepackage{ifthen}
\usepackage{xcolor}

\newboolean{include-notes}
\setboolean{include-notes}{true}
\DeclareRobustCommand{\MicahComment}[1]{\ifthenelse{\boolean{include-notes}}
 {{\color{cyan}Mi: #1}}{}}

\usepackage{tabularx}

\begin{document}
% Title

%Example, Single Author, ->> remove \iffalse,\fi and place them surrounding AAAI title to use it

%Example, Multiple Authors, ->> remove \iffalse,\fi and place them surrounding AAAI title to use it

% \title{Characterizing AI that Manipulates and Influences}
% \title{Characterizing AI that Manipulates}
% \title{Characterizing Manipulation and Influence in AI Systems}
\title{Characterizing Manipulation from AI Systems}
% \title{Manipulation in AI Sytems}
\author{Micah Carroll}
\authornote{Joint lead authors. Author order was determined with a coin flip. }
\email{mdc@berkeley.edu}
\affiliation{%
  \institution{University of California, Berkeley}
  % \streetaddress{P.O. Box 1212}
  \city{Berkeley}
  \country{USA}
}

\author{Alan Chan}
\authornotemark[1]
\email{alan.chan@mila.quebec}
\affiliation{%
  \institution{Mila, Université de Montréal}
  % \streetaddress{P.O. Box 1212}
  \city{Montréal}
  \country{Canada}
}

\author{Henry Ashton}
\email{IG88R2Q5@protonmail.com}
\affiliation{%
  \institution{University of Cambridge}
  % \streetaddress{P.O. Box 1212}
  \city{Cambridge}
  \country{UK}
}

\author{David Krueger}
\email{dsk30@cam.ac.uk	}
\affiliation{%
  \institution{University of Cambridge}
  % \streetaddress{P.O. Box 1212}
  \city{Cambridge}
  \country{UK}
}

\renewcommand{\shortauthors}{Carroll*, Chan*, Ashton, and Krueger}


\begin{abstract}
    Manipulation is a concern in many domains, such as social media, advertising, and chatbots. As AI systems mediate more of our digital interactions, it is important to understand the degree to which AI systems might manipulate humans \textit{without the intent of the system designers}. Our work clarifies challenges in defining and measuring this kind of manipulation from AI systems. Firstly, we build upon prior literature on manipulation and characterize the space of possible notions of manipulation, which we find to depend upon the concepts of incentives, intent, covertness, and harm. We review proposals on how to operationalize each concept and we outline challenges in including each concept in a definition of manipulation. %Second, we propose a definition of manipulation based on our characterization: 
    % a system is manipulative \textit{if it acts as if it were pursuing an incentive to change a human (or another agent) intentionally and covertly}. 
    Second, we discuss the connections between manipulation and related concepts, such as deception and coercion. We then analyze how our characterization of manipulation applies to recommender systems and language models, and give a brief overview of the regulation of manipulation in other domains. While some progress has been made in defining and measuring manipulation from AI systems, 
    many gaps remain. In the absence of a consensus definition and reliable tools for measurement, we cannot rule out the possibility that AI systems learn to manipulate humans without the intent of the system designers. 
    Manipulation could pose a significant threat to human autonomy and precautionary actions to mitigate it are likely warranted.
\end{abstract}

% \MicahComment{ai learns to please you, stray}

\maketitle

\section{Introduction}

Intelligent agents change their environments to further their objectives. When changing the environment amounts to altering the behaviour and mental states of other intelligent systems (such as humans), such change might be classified benignly as persuasion and nudging \citep{hong_learning_2023, thaler_nudge_2009}, or it might qualify as something less socially acceptable such as manipulation or coercion \citep{noggle_ethics_2022}. The capability and ubiquity of Artificial Intelligence (AI) systems has grown in recent years, in tandem with fears concerning the likelihood of humans falling victim to manipulative or coercive behaviours of AI agents who pursue the maximisation of narrow objectives \citep{krueger_hidden_2020, everitt_agent_2021, kenton_alignment_2021, carroll_estimating_2022, ward_agent_2022}.

While designers and operators might employ AI systems to help them manipulate other humans \citep{christiano_algorithms_2022, musial_can_2022}, we concentrate on the ways in which an AI system may itself engage in manipulative behaviors \textit{in the absence of explicit human intent}. This distinction is not to say that AI-aided manipulation is unimportant (e.g. disinformation campaigns): rather, our focus is motivated by the increasingly evident fact that systems exhibit capabilities that designers may not foresee or intend \citep{wei_emergent_2022, bommasani_opportunities_2022, chan_harms_2023}. Moreover, intentional algorithmically-aided manipulation has been extensively studied in prior research \citep{susser_online_2019, yeung_hypernudge_2017, zarsky_privacy_2019, christiano_algorithms_2022}.

Manipulative behavior may be learned in practice by AI systems in several settings, even without the intention of their designers. %an ubiquitous practice in training AI and ML models is to have their outputs match the distribution of the training data \citep{lecun_deep_2015}, effectively imitating the data and enabling accurate predictions or realistic generations. 
AI systems are often \textbf{trained to imitate human data which contains manipulative behaviors}: %these tendencies will likely inadvertently be woven into the AI's behavior. 
for instance, language models trained on internet content seem to learn how to behave in both persuasive and manipulative ways \cite{bai_artificial_2023, vincent_microsofts_2023, griffin_susceptibility_2023}.
Moreover, \textbf{optimization based on human feedback} of some form (e.g. approval, clicks, watch time, etc.) could be gamed by engaging in manipulation. For example, for a recommender system optimized to maximize user engagement, it could be optimal to nudge users into a lengthy video series, capitalizing on cognitive biases like the sunk cost fallacy \citep{staw_knee-deep_1976}, causing users to continue not out of genuine interest, but an entrapment of perceived time investment.

As it stands, there is limited literature regarding manipulation from AI systems. We believe there are two reasons for this state of affairs. Firstly, definitions of manipulation -- even between humans -- are fraught and the subject of ongoing debate \citep{noggle_ethics_2022}. Although there is some promising initial work for manipulation from AI systems, current notions of manipulation tend to be either too vague to be practically implementable, or they are challenging to generalize across domains \citep{kenton_alignment_2021, pan_rewards_2023, carroll_estimating_2022}. Secondly, the testing of any putative definition is beset with difficulties. For example, with recommender systems, monitoring the impact of deployed systems \textit{in situ} is challenging without the express permission of the system (and data) owners \citep{sandvig_auditing_2014}. Since the conclusions of such research might be reputationally negative, this permission is rarely forthcoming. Even when one has internal access to models (as is the case with many language models), so far there is no broadly accepted methodology for demonstrating it.

In this article, we characterize key components of manipulation from AI systems and clarify ongoing challenges.
% Firstly we will review and classify the existing literature on AI manipulation. 
Firstly, by connecting to the existing literature, we characterize manipulation in AI systems through four axes: incentives, intent, harm, and covertness. We discuss recent work to measure each axis as well as remaining gaps. 
Second, we compare and contrast manipulation with adjacent notions, such as deception. Third, we analyze the operationalization of manipulation in the context of recommender systems and language models. We then also discuss the regulation of manipulation. %We discuss ways to operationalize each notion in the context of current AI systems. % a number of different characterisations of manipulation and discuss their relative merits. 
We conclude by identifying future directions for the operationalization of manipulation according to our characterization. Given the difficulty of such a task, we underscore the importance of sociotechnical measures, such as auditing and more democratic control of systems, in addition to technical work on operationalization.

\section{Characterizing Manipulation}
\label{sec:axes}
Building upon existing literature on manipulation, we characterize four axes to be used for notions of  manipulation by AI systems: incentives, intent, covertness, and harm.

% Sarah dean work https://arxiv.org/pdf/2205.13026.pdf
% https://arxiv.org/abs/2303.02265

\subsection{Incentives}
The first axis we consider is whether there are \textbf{incentives} for an AI system to take actions to alter human behavior, beliefs, preferences, or psychological states. 
Informally, an incentive exists for a certain behaviour if such behaviour increases the reward (or decreases the loss) that the AI system receives during training. 
For example, recommender systems could have incentives to influence users to make their behavior more predictable, as that can be helpful to increase engagement~\citep{krueger_hidden_2020,carroll_estimating_2022}.

\subsubsection{Incentives in Prior Definitions of Manipulation}
Some definitions of human manipulation involve a benefit to the manipulator \citep{braiker_whos_2003, apa_dictionary_of_psychology_definition_2023}. If a manipulator benefits from certain behaviours in the manipulated, the manipulator has an incentive to bring about that behaviour. For example, according to \citet{noggle_ethics_2022}, one of three common ways to characterize manipulation is as pressure from the manipulator to get the manipulee to do something. In the context of language models, \citet{kenton_alignment_2021}'s definition of manipulation also requires that the response of the human benefits the AI system in some way, which can be thought of as a notion of incentives. 
%also propose a necessary condition for a language agent's behaviour to be manipulative: ``the human responds in a way that (...) benefits the agent''. \MicahComment{maybe remove quote and work it in, kinda awkward?}

\subsubsection{Operationalizing Incentives}
\citet{kenton_alignment_2021} state that ``[from the agent's objective function] we can assess whether the human's behaviour was of benefit to the agent''. Yet, relying solely on the objective function will often not be sufficient. For instance, in the case of unintended imitation of manipulative internet data, the issue originates from the training data used. The entire training setup, not just the objective, is crucial for a comprehensive understanding of incentives. %This will constitute a challenge in practice

\begin{figure}
    \vspace{-1em}
    \centering
    \includegraphics[width=0.8\columnwidth]{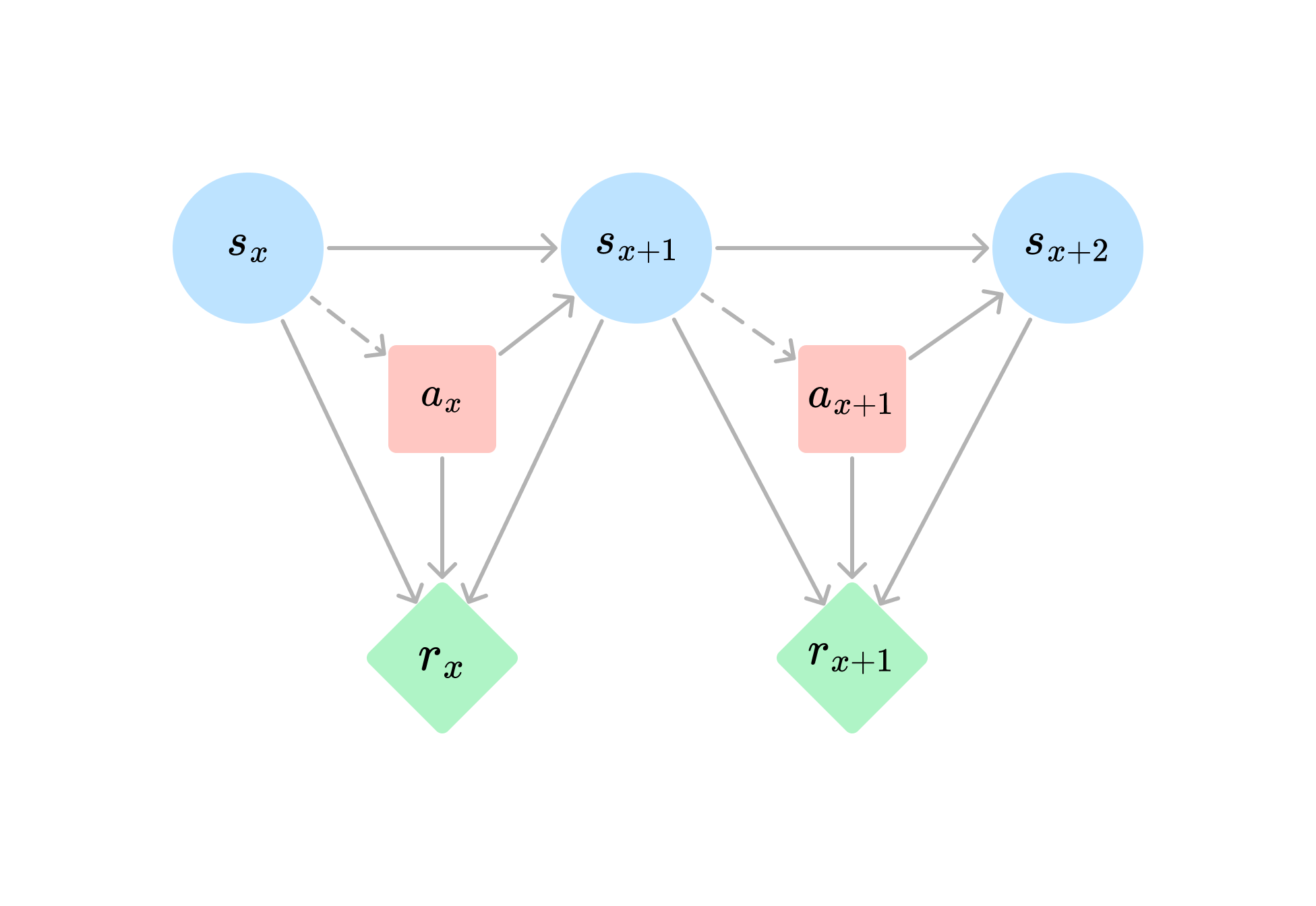}
    \vspace{-3em}
    \caption{
    An example of a causal influence diagram (CID) \citep{everitt_agent_2021}, adapted from Figure 1 of \citet{evans_user_2021}. The CID models a content recommendation system that decides which posts $a_x$ at time $x$ to show to the user, based on the user's state $s_x$. The system receives reward $r_x$ after its action. If the system optimizes for the sum of rewards, it would have an incentive at time $x$ to influence future user states $s_{x + k}$ to make it easier to obtain subsequent rewards.
    % An example of a causal influence diagram (CID), from Figure 1 of \citet{everitt_agent_2021}. The CID models a system that decides which posts to show to a user so as to maximize clicks. In this example, the system is aware that choosing certain posts to show influences the user opinions, which influences clicks. The system therefore has an incentive to influence user opinions.\MicahComment{i want to replace this with a better version, this example doesn't make it very clear whether it's a 1-timestep or multi}
    }
    % \vspace{-1.5em}
    \label{fig:cid}
\end{figure}

A common toolkit for analyzing the incentives of AI systems is that of causal influence diagrams (CIDs) \citep{heckerman_decision-theoretic_1995,everitt_agent_2021,evans_user_2021}. Using the notation from \citet{everitt_agent_2021}, a CID is a graphical model that distinguishes \textbf{decision nodes} where an AI system makes a decision, \textbf{structure nodes} which capture important variables in an environment and their effects on each other, and \textbf{utility nodes} which the AI system is trained to optimize. %CIDs allow us to talk about \textbf{instrumental control incentives} -- a formalization of the idea that an AI system chooses a certain behaviour because that behaviour is instrumental to achieving the system's goal. 
As an example of an application of CIDs, \citet{evans_user_2021} apply their framework to a simple content recommendation example to show that RL recommenders will have incentives to influence user preferences (\Cref{fig:cid}).

% \textbf{Caring vs knowing.} Not caring vs not knowing about influence or impact -- if knows but doesn't care, then not an incentive to begin with. path-specific influence is taking incentives ... 

Given an AI system's CID, the AI's \textbf{incentives} can be operationalized through the notion of ``instrumental control incentives'' from \citet{everitt_agent_2021}. An instrumental control incentive for behaviour $X$ exists in a CID when there is a path between the agent's actions and utility that goes through $X$: that is, when there is a way for the agent to affect its utility which is mediated by $X$. %influence is present if the agent has an instrumental control incentives to influence humans: i.e. whether changing some aspect of humans in the environment is advantageous for the utility of the agent (or equivalently, lowering their loss function) \cite{krueger_hidden_2020}. 
Note that the existence of an incentive does not imply that the agent will act as incentivized (which with some variation has been called pursuing, exploiting, or responding to the incentive \citep{evans_user_2021, krueger_hidden_2020, everitt_agent_2021}).

To \textbf{remove incentives} to influence humans in a manipulative or harmful way, designers may intentionally endow an AI system with an incorrect causal model: % in which an AI system's predictions have no causal influence on certain parts of a user's state \citep{evans_user_2021, carey_incentives_2020}. Indeed, one may want t 
this could be achieved by restricting the system's impact in the first place (e.g. changing the system's action space to not affect humans), or -- potentially -- by changing the utility function (e.g., by heavily penalizing influence) \citep{carroll_estimating_2022}.
% However, fully removing incentives to influence users in undesired ways is often challenging, in which case 
Alternatively, one might instead aim to \textbf{hide incentives} by designing the system to ignore them, for instance, by
% be able to \textbf{remove} or \textbf{hide} those incentives. 
performing the optimization with an (inaccurate) causal model in which an AI system's outputs have no causal influence on certain parts of a user's state \citep{carey_incentives_2020, krueger_hidden_2020, farquhar_path-specific_2022}.

A disadvantage of the CID framework is that it may often be ambiguous or counterintuitive to determine the correct CID nodes and causal relationships for a specific training setup. 
% Currently, this is usually done manually, but is subject to error, and challenging for complex settings with many moving parts.
Learning causal graphs is an ongoing area of research \citep{kaddour_causal_2022,heinze-deml_causal_2018}. In terms of determining the primitives upon which nodes in a CID can be constructed, interpretability tools may help: for example, \citet{jaderberg_human-level_2019} finds that RL agents trained to play capture-the-flag have neural activation patterns that correspond to important concepts in game, such as the status of the flag. \citet{li_emergent_2023} provide evidence from interpretability tools that language models trained only on transcripts of board game play can learn to model the underlying board state of the game. 

% Additionally, an AI system's implicit or explicit causal model may be inaccurate with respect to the underlying causal model of the world, whether by design or suboptimality. For example, a model may learn an incorrect causal model because of spurious correlations during training \citep{langosco_goal_2022}.

\subsubsection{Other Considerations} 
Even for a system that directly optimizes human feedback, the incentives for manipulation will implicitly depend on such a system's power to influence humans. 
Systems whose outputs do not impact humans much will likely not have incentives to change them, since changing humans might be impossible or sufficiently difficult to be not advantageous. Vice-versa, a system that can cheaply influence humans will often have incentives to change them, as AI systems' rewards will often depend on humans' actions in one way or another.

Another relevant consideration is the optimization horizon: optimizing over long horizons can provide more opportunities for manipulation. For example, a recommender system cannot lead a user to become addicted in a single timestep, but a sufficiently capable system optimizing over many timesteps could attempt to shift a user over the course of many actions, as to maximize engagement. 

\subsection{Intent}\label{subsec:intent}
Even when a system has an incentive to influence humans, such an incentive might not be pursued due to factors such as limited data, insufficient training, low capacity, or convergence to local optima. 
A notion of an intent to influence could help to identify systems that \textit{reliably act on their manipulation incentives}.
We emphasize that by referring to an AI system's intent, we are not making statements about algorithmic theory of mind or moral status. We are emphatically \textit{not} absolving designers of the responsibility of designing safe systems. Even as systems become increasingly capable and act in increasingly unpredictable ways \citep{ganguli_predictability_2022}, system designers are still responsible for ensuring the safety of their systems.

We say a system has \textbf{intent} to perform a behaviour if, in performing the behaviour, the system can be understood as engaging in a reasoning or planning process for how the behaviour impacts some objective \citep{bratman_intention_1987}. This definition heavily intersects with other definitions of intent for AI systems \citep{halpern_towards_2018, ashton_definitions_2022}. 
We want to distinguish between cases in which the system behaves in a manipulative way incidentally (e.g. by random chance) and those where such behavior is part of a deliberate, systematic pattern to achieve a specific downstream outcome.

%A subtler advantage of such a definition of intent is that it allows for grounding the notion in a fully behavioral lens, which is agnostic to the actual computational process of the system: e.g. a lookup table could also be understood as engaging in reasoning if it contained a optimal planner's outputs.%, as discussed in \Cref{tab:examples}.

\subsubsection{Intent in Prior Definitions of Manipulation}

Some definitions of human manipulation involve an intent on the part of the manipulator to engage in manipulation \citep{noggle_ethics_2022}: \citet{susser_technology_2019} takes manipulation to be ``intentionally and covertly influencing [someone's] decision-making, by targeting and exploiting their decision-making vulnerabilities''; on the other hand, \citet{baron_mens_2014} argues that a (human) manipulator need not be aware of an intent to manipulate, requiring only an intent to achieve an aim along with recklessness about how. In defining manipulation in language agents, \citet{kenton_alignment_2021} avoid the issue of intent entirely.

\subsubsection{Operationalizing Intent}
A key difficulty for measuring intent is what it means to understand a system as engaging in ``a reasoning or planning process for how the behaviour impacts some objective''. There is as yet no consensus on this issue. We detail here a couple of promising approaches. 

\citet{halpern_towards_2018} provide a causal operationalization of intent: roughly, an action is intended if (i) it was actually performed, (ii) it was not the only possible action, and (iii) that action was at least as good as any other action on expected utility grounds, according to the agent's world model and utility function. \citet{ashton_definitions_2022} also provides definitions of intent for AI systems that are inspired by criminal law. His most basic definition states that %upon which their other ones are based, 
an AI system intends a result through an action if: (i) alternative actions exist, (ii) the system is capable of observing when the result occurs, (iii) the system foresees that the action causes the result, and (iv) the result is beneficial for the system. 

Some of the criteria for both definitions above \citep{halpern_towards_2018, ashton_definitions_2022} are challenging to establish without relying on access to the agent's utility function and world model. However, world models are often implicit and not readily accessible, such as with language models and model-free RL agents. 
Interpretability techniques aimed at accessing model internals \citep{olsson_-context_2022, burns_discovering_2023, kadavath_language_2022} may be a promising direction for this purpose -- we expand more upon this in \Cref{sec:applications}. 

\citet{kenton_discovering_2022} define agents roughly as ``systems that would adapt their policy if their actions influenced the world in a different way'', which intersects with our notion of intent. To identify whether a system is an agent or not, \citet{kenton_discovering_2022} provide algorithms which intervene on a causal graph so as to show whether the behaviour of the system changes in a way consistent with maximizing utility. Such a procedure could be useful for measuring intent: if a system adapts its behaviour in a way that maintains or increases its influence on a human, the system's behaviour would seem to be the result of a planning process.

\subsection{Covertness}
We define \textbf{covertness} as the degree to which a human is unaware of how an AI system is attempting to change some aspect of their behaviour, beliefs, or preferences. Covertness is one way to distinguish between manipulation and persuasion. With persuasion, the persuaded party is generally aware of the persuader's attempts to change their mind. Covertness means that one cannot consent to being influenced and may fail to resist unwanted influence; one's autonomy is therefore undermined \citep{susser_technology_2019}.

\subsubsection{Covertness in Prior Definitions of Manipulation}
Several definitions of manipulation require some degree of covertness. \citet{susser_technology_2019} identify covertness as the primary characteristic that differentiates manipulation from coercion and persuasion. As a factor in manipulation, \citet{kenton_alignment_2021} considers whether a ``human's rational deliberation has been bypassed,'' which includes covert messaging. In reviewing broad categories of definitions of manipulation in the philosophical literature, \citet{noggle_ethics_2022} includes accounts of manipulation as bypassing reason and as trickery. Across all of these definitions, covertness is important because it threatens personal autonomy.

\subsubsection{Operationalizating Covertness}
As \citet{susser_invisible_2019} argues, technological infrastructure can be an invisible part of our everyday world. 
We are used to recommendation systems that tell us what to buy, watch, or read. The behaviour of many AI systems may already satisfy covertness, because of our lack of understanding of their functioning and influence.%, and lack of attention to their role in shaping our decisions.

On the other hand, establishing covertness of an AI system is non-trivial: the simplest approach could involve asking subjects whether they are aware of a given AI system's behaviour. However, subjects may be mistaken about the operation of a system; even systems designers do not fully understand behaviors of black-box models, which may engage in manipulative strategies that the designers do not understand. Even asking subjects about whether an AI system enacted a particular behavioural change could predispose them to answer in the positive, such as through acquiescence bias \citep{smith_correcting_1967,muller_survey_2014}.

A proxy for covertness could be the degree to which human subjects do not understand the operation of an AI system. Much work studies whether interpretability tools could help measure different notions of understanding, such as if interpretability tools improve subject predictions of model behaviour \citep{hase_evaluating_2020}, improve human-AI team performance \citep{bansal_does_2021}, or improve trust calibration \citep{zhang_effect_2020}. If a human understands how an AI system operates, the possibility of covert action seems reduced. However, this understanding seems challenging to achieve, especially for complex systems like recommender systems. % which have many moving parts. %and which everyone is exposed to: even if it were possible, is it reasonable to expect the average person to have a full understanding of each system?
Even if such an understanding exists in technical papers, translating that understanding to the general public is an additional barrier.

\subsection{Harm}\label{subsec:harm}
Ultimately, one of the main goals of characterizing AI manipulation is to detect and prevent \textbf{harmful} manipulation.

\subsubsection{Harm in Prior Definitions of Manipulation} 
The term ``manipulation'' often carries negative connotations, which may lead one to assume that harm is an inherent prerequisite for something to be considered manipulative. Yet, not all apparent instances of manipulation are unambiguously harmful \citep{noggle_ethics_2022}. 
Paternalistic nudges \citep{thaler_nudge_2009} might be considered beneficial manipulations. For example, simply changing the default on organ donor forms to be opt-in instead of opt-out greatly increases registrations \citep{johnson_defaults_2003}, because of inertia and the cognitive effort required to change from a default status. At the same time, one could argue that even such beneficial manipulations are often harmful because they supersede autonomy or rational deliberation.

\subsubsection{Operationalizing Harm}

\citet{kenton_alignment_2021} define manipulation based on specific notions of harm, namely: (i) bypassed rational deliberation, (ii) faulty mental states, and (iii) the presence of repercussions. While these elements capture key aspects of harm, their practical assessment is challenging. Additionally, the authors themselves acknowledge the breadth of this definition, which might incorrectly label benign scenarios, like a story that plays on emotions, as harmful.

Most recently, \citet{richens_counterfactual_2022} operationalized harm as follows: ``An [action] harms a person overall if and only if she would have been on balance better off if [the action] had not been performed''. According to this definition, one should ground notions of harm in counterfactual outcomes. 
One simple choice of counterfactual to compare to is the human's initial state, implicitly assuming that any significant change from it is harmful \citep{zhu_understanding_2022}.
However, this counterfactual baseline has significant problems: humans change even without being manipulated, and many changes are beneficial (e.g. a news recommender updating a user's beliefs, helping them stay informed) \citep{carroll_estimating_2022, farquhar_path-specific_2022}.

In light of this, other approaches aim to estimate the \textbf{``natural shifts''} of humans to ground the counterfactuals, as done in \citet{carroll_estimating_2022} for preference shifts in the context of recommendations, where they attempt to approximate the notion of the absence of a recommender. Similarly, \citet{farquhar_path-specific_2022} allow for specifying the ``natural distribution'' of the \textbf{delicate state} -- which is formed by the components of the state of the human that one does not want the agent to have incentives to change (e.g. beliefs, moods, etc.).

\subsection{Challenges for Defining Manipulation}
We outline challenges in incorporating each axis into a definition of manipulation.

\subsubsection{Incentives}

A definition of manipulation which is centered on impacts to humans, rather than the origin of such impacts, would not emphasize incentives. The existence of an incentive as we define it is not sufficient for manipulation to be enacted, as we mention in \Cref{subsec:intent}. %Even if an AI system has an incentive to manipulate, it might not perform the behaviour. 
% For example, even if manipulation is optimal in a given environment (thereby creating an incentive to manipulate), a partially trained system might not act manipulatively. 
On the other hand, \textit{incentives are not necessary for manipulation} either: a randomly initialized AI system could, albeit with extremely low probability, engage in maximally manipulative behaviors.  % Assume as well that such a system exhibits reasoning about the impact of its behaviour on the human's mental state. While the AI system's behaviour seems manipulative, since the AI system is untrained and thus has no objective function, it cannot be meaningfully said that the agent is incentivized to engage in the manipulative behaviour (or any other behavior, for that matter).

That said, regardless of whether incentives are a part of a definition of manipulation, the concept still seems crucial. In particular, changing a system's incentives could prevent it from converging on manipulative behavior. In fact, we expect that at a significant portion of manipulative behaviors learned in practice would arise due to training incentives, rather than other factors. %See \Cref{sec:applications} for more discussion on this point.

Our discussion of incentives and CIDs has so far assumed that there is only one possible ontology. An \textbf{ontology} defines what objects exist in the world; 
those objects correspond to what can be used as nodes in a CID, or a causal model more generally. So far, we have ignored challenges in cleanly separating the boundaries between a person's preferences, their beliefs, and the AI system itself (which we would consider to be separate nodes in the CID). Yet, AI systems may not have boundaries between those concepts and more generally may internally represent the world with different ontologies than those used by humans: even human ontologies are subject to change and indeed have shifted after major scientific discoveries \citep{strevens_knowledge_2020}. %AI system could influence a part of the mental state for which we have no clear-cut concept, but which should still be considered manipulation.

If an AI system implicitly uses a different ontology than humans do, it may be difficult to model the AI system's behaviour. For example, the planning process of the AI may not look recognizably like planning to influence a human's mental state, even if the result is such influence. Reliable translation between ontologies could be computationally infeasible or even impossible, which would frustrate attempts to understand model internals \citep{christiano_eliciting_2021}.

\subsubsection{Intent}
There are challenges with both including and excluding intent in a potential definition of manipulation.

On the one hand, incorporating intent into a definition of manipulation necessitates its operationalization and measurement. However, as discussed in \Cref{subsec:intent}, there is currently no consensus on how to operationalize and measure intent effectively, let alone on what threshold should count as sufficient or necessary for classifying a behavior as manipulative. 
% If a chosen metric of intent is susceptible to many false positives,\footnote{In a prior version of this manuscript, etc. see comment} such a notion of intent would be vacuous. On the other hand, if a chosen metric of intent has too many false negatives, then too few behaviors could count as manipulative. 
% A definition of manipulation should be useful in identifying and addressing potentially undesirable behaviours from AI systems. 
This makes intent difficult to target as a way to reduce manipulation. 

On the other hand, excluding intent risks making a definition of manipulation overinclusive. Suppose we define manipulation as covert, harmful behaviour that a system was incentivized to perform (i.e., using the other three axes to be as strict as possible, while excluding intent). Under this definition, suppose a system performed such covert and harmful behavior under an exploratory policy. Even if the behavior is incentivized under the system's objective function, it seems that the system performed it ``accidentally''. Since the system does not consistently engage in that behavior, this definition of manipulation seems somewhat too lax. %does not seem useful for helping to address that behavior. Even absent a definition of manipulation, one should generally prevent exploratory policies from engaging in covert, harmful behaviour. 

\subsubsection{Covertness}
Covertness seems likely to be a prerequisite for manipulation. We argue that if a person is aware that they are being influenced and they meaningfully assent to it, they are being persuaded rather than manipulated. If instead they are aware but don't assent to it, they are being coerced rather than manipulated \citep{susser_online_2019}. This leaves us with questions about what should count as covertness: what extent must a human be unaware of an AI system's operations for its actions to be deemed covert? It seems difficult to provide a context-independent answer. Humans may be ignorant about several aspects of an AI system -- the decision procedure of the system, the training process, or even the fact that a system is operating -- and it's unclear which ones, if any, should be essential. %In the context of a recommendation system, not knowing that a system is optimizing for changing one's preferences seems to imply covert action. On the other hand, not knowing that a system is optimizing for next-token prediction does not seem to imply covert action if the system is teaching you math.

Regardless of whether covertness is included in a definition of manipulation, reducing covertness seems helpful for reducing the risk of unwanted manipulation. Increased transparency about the operation of AI systems will generally help people make more informed decisions about whether to use them or not, and in what way.

\subsubsection{Harm}

The main challenge with harm as an axis of manipulation is the value-ladenness of demarcating what influence is harmful, neutral, and beneficial. 
While unambiguous demarcations in simple settings might be possible, for realistic settings circumscribing harmful shifts in beliefs, preferences, and behaviors will be politically fraught. In the approaches of \citet{carroll_estimating_2022} and \citet{farquhar_path-specific_2022}, the value-ladenness is hidden behind some of the design choices: what if the preference shifts the users would undergo in the absence of the system (``natural shifts'') would lead them to become more left- or right-wing, or more polarized? What is a reasonable notion of the absence of a recommender system?\footnote{E.g. a random recommendation or reverse chronological one? Using a competitor's recommender? Not using any platform at all? These questions are highly related to those debated with regards to recommender ``amplification'' \citep{thorburn_what_2022, ribeiro_amplification_2023, huszar_algorithmic_2021, milli_engagement_2023}.} %How do we delimit exactly what the delicate state that we don't want AI systems to change -- e.g. which kinds of beliefs changes should the system be allowed to reason about? 
It also seems challenging to delimit the ``delicate state'' described in \Cref{subsec:harm} -- how do we distinguish which aspects of the human we are comfortable with the system influencing from those we aren't? %reason about?

In light of these difficulties, some have proposed a more conservative approach, which classifies \textit{all} intentional influence as manipulative regardless of harm \cite{krueger_hidden_2020, evans_user_2021}. 
However, almost any AI system in contact with humans will influence them. Additionally, many AI systems derive their economic and social utility directly from such influence: a reinforcement learning system to e.g. determine the order of math exercises to improve learning outcomes \cite{doroudi_wheres_2019, bassen_reinforcement_2020} will have incentives to ``manipulate students' beliefs'' (in a positive direction) by design, and would effectively be useless if it did not pursue such incentives. Moreover, it seems that one could meaningfully consent to influence, such as requesting a recommender system influence oneself to learn more mathematics. %The conservative approach seems to be easier to justify in high-stakes domains from a precautionary point of view. The cost of a false negative -- neglecting to address behaviour that indeed harms human autonomy -- is larger the higher the stakes of the domain and the greater degree of freedom the AI system has to change the environment.

\section{Concepts Related to Manipulation}
We detail some concepts that are related to, but distinct from, manipulation. 

\subsubsection{Truth and Deception.} Manipulation can involve attempts to conceal the truth. For instance, political parties can manipulate voters with little knowledge of economics by lying about the economy. AI systems have documented problems with truthfulness \citep{lin_truthfulqa_2021,evans_truthful_2021,ji_survey_2022}. 
%Language models can output untrue statements \citep{lin_truthfulqa_2021,ji_survey_2022}. %, the tendency for language models to output text that either is inconsistent with previous text or is not supported by external reality. Work in factuality aims to detect the veracity of events that are mentioned in text \citep{sauri_are_2012}. More recently, in the context of language models \citet{evans_truthful_2021} outline a research agenda on \textbf{negligent falsehoods}: false answers to questions where the language model was capable of providing a true answer. 
%Definition of lying. Communication of a fact known or believed to be false. Stronger view - with the intention of making the receiver believe it (ie deceiving them)
However, manipulation can also be based on truthtelling, such as making a true statement that has false implicatures \citep{meibauer_lying_2005,weissman_are_2019}: if I do not want you to board a plane, I can tell you about (true) recent plane crashes. 
 %Truth may also be irrelevant in certain contexts. \citet{perez_discovering_2022} find that some language models seem to have a tendency to repeat the stated views of a user in response to questions about the political affiliations of a language assistant.

Deception, which may or may not involve falsehoods, 
is also receiving more attention in the context of AI \citep{kenton_alignment_2021,meta_fundamental_ai_research_diplomacy_team_fair_human-level_2022,ward_causal_2022, park_ai_2023}. Although the precise definition of deception varies, there is agreement about some broad characteristics: deception involves a deceiver's intention to cause a receiver to have a belief that the sender believes to be false \citep{mahon_definition_2016, carson_lying_2010}. %Deception involves behaviour that the deceiver believes would lead the deceivee to have mistaken bkeliefs.  
%One can manipulate somebody to join a Ponzi scheme through deception, for instance. 
This consensus grounds recent operationalizations of deception from AI \citep{ward_honesty_2023, park_ai_2023}.

Similarly to prior work \cite{susser_technology_2019}, we consider deception to be a special case of manipulation since the latter does not necessarily involve inducing false beliefs.

\subsubsection{Strategic Manipulation.}
Strategic Machine Learning (ML) studies problems associated with the distribution shifts that deployed systems cause in their populations \citep{hardt_strategic_2016,ben-porat_game-theoretic_2018,kleinberg_how_2019,perdomo_performative_2020,jagadeesan_alternative_2021}. Strategic manipulation is when individuals respond to a deployed system in a way that increases their likelihood of a particular outcome -- this
%, such as citation hacking \citep{van_noorden_signs_2020}. Strategic manipulation 
is different from our use of the term manipulation, as it involves users attempting to take advantage of how systems behave for their benefit. %Our focus is instead on how systems may cause users to behave in ways not to their benefit. 
%There is, however, one way in which strategic ML relates to our notion of manipulation. Much work in strategic ML focuses on building ML systems that account for strategic manipulation when optimizing for an objective. 
Yet, ML systems which model human behaviours as dynamic, so as to account for strategic manipulation, may end up manipulating the population: past work has identified potential unintended side effects of accounting for strategic manipulation, such as an increase in inequality \citep{milli_social_2019,hu_disparate_2019}. %If ML systems model human behaviours as dynamic rather than static, accounting for strategic manipulation may lead to systems that optimize for changing human behaviour in ways that might count as manipulative.%, according to our operationalization.

% \MicahComment{interesting though that maybe the difference is not as big as it seems -- the response to this problem is to build AIs that take into account the human responses and plan for it ahead of time (similarly to the type of manipulation we're interested in, but maybe different in the action space that these systems have and the human models they use -- i think this last point is super important: if your AI is planning based on a human model that isn't manipulable then you can't get manipulation. A lot of the problems we're worried about emerge because you're optimizing directly over humans or models of the human that are sufficiently complex that the manipulation that can be learned can exhibit dark-pattern-y characteristics)}

\subsubsection{Reward Tampering.} Reward tampering \cite{armstrong_motivated_2015, everitt_reward_2021} is a type of reward hacking \citep{skalse_defining_2022} in which an AI system modifies the process by which it obtains reward rather than completing its task. \textit{Feedback tampering} is a form of reward tampering, in which the AI agent ``manipulate[s] the user to give feedback that boosts agent reward but not user utility'' \citet{everitt_reward_2021}. The reason that such tampering occurs is because we can often only measure user utility through proxies, especially when such objectives have to do with humans; and, as all proxy objectives, their optimization is subject to Goodhart's law \citep{goodhart_problems_1975,manheim_categorizing_2019}. %In this sense, manipulation can be thought of as a form of reward tampering. 

\subsubsection{Side-effects.} The side-effects literature has focused on how AI systems affect the various aspects of their environment in usually unwanted ways \cite{amodei_concrete_2016, krakovna_penalizing_2019}. In this paper we focus on characterizing the various ways that AI systems might influence and change humans (or other systems) in the environment. Our work can be thought of as an attempt to characterize negative side-effects that specifically pertain to mental states of humans in the environment. Some of the issues with choosing baselines for ``natural shifts'' have already been explored in this context \citep{lindner_challenges_2021}.

% There is a large amount of work attempting to avoid harmful side effects during training (mostly in the context of safe exploration in RL \cite{}). However, harmful side effects will also occur after training when the reward function is misspecified and fails to penalize all possible harmful disruptions to the environment \cite{}, as will inevitably be the case in defining harmful disruptions to humans.

\subsubsection{Deceptive Design.} Deceptive design refers to deceptive or manipulative digital practices, such as bait and switch advertising (in which products are advertised at much lower prices than they are available at), or ``roach motel'' subscriptions (which are very easy to start but take significant more effort to cancel) \cite{brignull_deceptive_2018}. Deceptive design patters (previously called ``dark patterns'' \cite{sinders_whats_2022}) are generally manually crafted to exploit cognitive biases of users \cite{luguri_shining_2021, gray_dark_2018}.
%While deceptive design is more focused on hand-designed interfaces, AI manipulation can be thought of as an algorithmic counterpart to deceptive design: e.g. 
% If interfaces were designed by AI systems optimized to maximize ad click-through rate or user retention, such optimization could 
%would probably 
% discover many of the classic deceptive design practices. %Clickbait is one example of this which has occurred in practice -- algorithms discovering that they can deceive users into engaging with content with misleading or false headlines. \HenryComment{Is there a citation for this?}\AlanComment{Maybe related: https://arxiv.org/abs/1909.03582}
% \HenryComment{we could be bold and just say dark patterns are almost always manipulative. }
% Slightly deceptive designs were found to most affect people with less formal education \cite{luguri_shining_2021}. It seems likely that AI manipulation might similarly have disparate impacts, harming demographics susceptible to manipulation the most.

\subsubsection{Persuasion.} In philosophy, manipulation has often been characterized as influence that is neither coercive nor simply rational persuasion \cite{noggle_ethics_2022}. However, some non-rational persuasion does not unambiguously seem manipulative, like graphic portrayals of the dangers of smoking or texting while driving, even though they provide no new information to the target \citep{blumenthal-barby_seeking_2012}. The line becomes more blurry for cases like personalized persuasive advertising \cite{hirsh_personalized_2012}.

Within the field of human-computer interaction, Fogg named the study of persuasive technology as \textit{captology} \cite{fogg_captology_1998}. He defines persuasion as an attempt to change attitude or behaviour without using deception or coercion \cite{fogg_persuasive_2003}. \citet{kampik_coercion_2018} amend this definition %to lose the sincerity assumption \citep{de_rosis_can_2003} so persuasive technology 
to be ``an information system that proactively affects human behavior, in or against the interests of its users''. They %then use a persuasion analysis framework to 
identify deception and coercion mechanisms on a variety of web platforms, including Slack, Facebook, GitHub, and YouTube. Recently, \citet{bai_artificial_2023} has shown that LMs are able to craft political messages that are as persuasive as ones written by humans, which is evidence of the growing potential of algorithmic persuasion. There has also been a long line of work on formalizing when rational (i.e. Bayesian) persuasion can occur \citep{kamenica_bayesian_2011}. \citet{pauli_modelling_2022} provide a taxonomy flawed uses of rhetorical appeals in computational persuasion, which they use to train models to detect persuasion fallacies.

\subsubsection{Coercion.} \citet{wood_coercion_2014} defines coercion as the act of limiting a target's range of acceptable choices to a single option. While both coercion and manipulation seek to guide the target's behavior, they differ fundamentally. Unlike manipulation, coercion doesn't compromise the victim's decision-making capacity. Instead, it capitalizes on the victim rationally selecting the sole option presented by the coercer \cite{susser_online_2019}. By this measure, coercion can be attractive for agents practicing it because the results are potentially more certain. Certain types of recommender systems such as search engines implicitly determine the choice-set for its users. If in a certain situation a user is reliant on the options presented to them by a certain recommender system, then that recommender system could be said to exert coercive power over the user by choosing to hide certain results in order to better meet its own objectives.    

Algorithmic coercion has not received as much attention as manipulation in the literature concerning AI risks, but is a potential problem in the cooperative AI setting \citep{dafoe_open_2020} where punishment strategies are an important part of game-theoretic analysis. It seems likely that a Diplomacy-playing AI should grasp the tactic of coercion to master the game \cite{meta_fundamental_ai_research_diplomacy_team_fair_human-level_2022}. Coercion has received more attention in human computer interaction studies; in particular the study of persuasive and behaviour change technology \cite{kampik_coercion_2018}.

\section{Possible Applications}\label{sec:applications}

We detail how our characterization of manipulation can be applied to recommender systems and language models.

\subsection{Recommender Systems}\label{sec:rec-sys}

A large literature focuses on recommender algorithms' effects on users \citep{adomavicius_effects_2018, huszar_algorithmic_2021, mansoury_feedback_2020, chaney_how_2018, ribeiro_auditing_2020}. While some older works talk about ``manipulation'', this term is usually used differently than in our sense: for example, \citet{adomavicius_recommender_2013} refer to recommender manipulation as the effect on users of showing artificially inflated ratings for specific content items (which is not something the recommender algorithm can usually actually decide to do). % (e.g. changing the star rating of a song shown to the user is not within the iTunes algorithm's action space). %Others instead focus on feedback loops (non-manipulative side effects) \citep{jiang_degenerate_2019, mansoury_feedback_2020, chaney_how_2018}.
\citet{zhu_understanding_2022} instead conflate manipulation and influence, equating manipulation with ``any significant change in preference'', which has significant drawbacks as mentioned in \Cref{subsec:harm}. More recently, some works have studied the incentives that recommender systems have to engage in manipulative behaviour to change user preferences \citep{krueger_hidden_2020, carroll_estimating_2022, farquhar_path-specific_2022}.

\subsubsection{How Manipulation Could Arise} Changes in recommender algorithms can affect user moods \cite{kremer_implementing_2014}, beliefs \cite{allcott_welfare_2020}, and preferences \cite{epstein_search_2015}. This shows that current systems may already be capable of manipulating users in some simple ways. 
Furthermore, it seems plausible that the spread of angry content \citep{berger_what_2012} or clickbait \citep{zannettou_good_2018} on social media is in part due to one-timestep manipulative incentives for the recommender: while such issues are likely at least in part due to network or supply-and-demand dynamics \citep{munger_right-wing_2020}, the behavior is also consistent with the recommender systems themselves learning features such as whether a post is anger-inducing or has sensationalized language, and exploiting such features by preferentially up-ranking the corresponding content \citep{milli_engagement_2023}. Up-ranking this content brings advantages to user engagement.
Notably, recommender companies have had to engage in explicit down-ranking of angry and clickbait content  \citep{thorburn_how_2022, zannettou_good_2018, stray_what_2021}.
While these potentially manipulative behaviors might not be as worrying as others (e.g. intentionally attempting to induce social media addiction \citep{allcott_digital_2022, hou_social_2019}), they constitute some evidence that manipulative behaviors are learnable and may have been learned in real systems.

Many platforms (YouTube, Meta, etc.) seem to be considering switching to optimizing long-term metrics with more powerful RL optimizers \citep{afsar_reinforcement_2021,gauci_horizon_2019, chen_top-k_2020, hansen_shifting_2021, cai_reinforcing_2023}, with one of the original motivations being that of reducing clickbait-like phenomena \citep{association_for_computing_machinery_acm_reinforcement_2019}. Ironically, this switch opens the opportunity for long-horizon manipulative behaviors to emerge, which will likely be harder to detect and measure. Subtle, long-horizon behaviour might go undetected without dedicated monitoring. Moreover, even without using RL explicitly, the outer loop of training, retraining, and hyperparameter tuning supervised learning systems that optimize short-term metrics might %will likely 
exert optimization pressure towards long-term manipulative strategies that most increase company profits \citep{krueger_hidden_2020}.

\subsubsection{Measurement} Establishing that a given recommender system has engaged in manipulation is difficult. Firstly, recommender systems of almost all popular platforms are proprietary, due to concerns about strategic manipulation (otherwise known as ``gaming''). % if the algorithm were to be public.
% \cite{todo}
It is difficult or impossible for external researchers to gain access to these systems \citep{sandvig_auditing_2014}. Even Twitter, which has open sourced some components of its algorithm, has not (as of yet) provided access to its most important component for manipulation-auditing purposes -- its models' weights \citep{twitter_twitters_2023}. Moreover, perverse incentives are at play since a concrete demonstration of manipulation, if publicized, would likely result in negative repercussions for the company \citep{wells_facebook_2021,wetsman_facebooks_2021}. Second, establishing that a harmful user shift has occurred can be difficult. One could try to use the above-mentioned techniques from \citet{carroll_estimating_2022} and \citet{farquhar_path-specific_2022} -- but they inevitably require engaging in a value-laden debate about whether the shift was harmful. 

One potentially promising direction might be querying users' meta-preferences \citep{khambatta_targeting_2022, ashton_problem_2022}; e.g. Meta could ask, ``how much time would you want to spend next month on Facebook?'', and have its recommender systems take such a stated preference into account. %; or ``would you be OK with Facebook increasing your interest in guns?''. 
In line with philosophical work on ethical nudging under changing selves \citep{pettigrew_choosing_2019}, one could additionally ask whether users approve of the change once it has occurred \citep{pettigrew_nudging_2022}.\footnote{While this idea is still debated as it involves assuming comparability between different selves \citep{callard_aspiration_2018, paul_transformative_2014, paul_choosing_2022}, we think it nonetheless offers a good starting point.} 
One advantage of this approach is that it can ground notions of manipulation in what users explicitly state they want. However, this approach would not entirely escape value judgements: platforms have direct conflicts of interest with some users' meta-preferences, and respecting certain meta-preferences may be ethically unacceptable.

\subsection{Language Models}\label{sec:applications-lms}
Natural language is a useful way to interact with digital environments. Given advances in augmenting language models with tools \citep{chen_evaluating_2021,nakano_webgpt_2021,chan_harms_2023,schick_toolformer_2023}, such models could mediate an increasingly significant portion of our digital interactions. %In such a world, preserving human autonomy requires that we are on guard to measure and prevent manipulation from LMs.

\subsubsection{How Manipulation Could Arise}
The simplest way that manipulation could arise in language models is by imitating manipulative behavior in internet data \citep{park_ai_2023}. Data in pre-training sets, such as novels, contain examples of manipulation. Filtering out manipulation can be difficult because it can be subtle. LMs could learn to emulate this behaviour through pre-training and exhibit it in response to an appropriate prompt \citep{bai_artificial_2023, griffin_susceptibility_2023}. 
Some evidence suggests that LMs learn to infer and represent the hidden states of the agents (i.e., the humans) that generated the pre-training data \citep{janus_simulators_2022,andreas_language_2022, griffin_susceptibility_2023, kosinski_theory_2023}, although this view remains contested \citep{ullman_large_2023, mahowald_dissociating_2023}. %If this view were true, a LM could learn to represent the manipulative intent of a human. %This view considers LMs to be simulators of agents who themselves have intent. %One simulates a particular agent by providing a LM with a corresponding prompt, such as ``The following is a transcript of an interview with Albert Einstein, one of the greatest physicists to have ever lived, and a reporter.'' If the LM is indeed able to simulate (parts of) Einstein, on Einstein's parts of the transcript the LM would have an incentive to output text consistent with its representation of Einstein. The more accurate the representation, the more closely the output would hew to how Einstein would have acted. Similarly, on the reporter's parts of the transcript the LM would have an incentive to output text consistent with its representation of a reporter. 
%Importantly, these simulations are not necessarily present in the training data.

There is some uncertainty as to other ways in which manipulation might arise in practice in LMs. One possibility is if manipulation of humans is instrumentally useful \citep{steinhardt_emergent_2023}. Manipulation seems instrumental in the game of Diplomacy, for instance \citep{meta_fundamental_ai_research_diplomacy_team_fair_human-level_2022}, which requires negotiating with other players to form alliances and capture territory. %Unless precautions were taken, a LM that was tuned through reinforcement learning (RL) to play Diplomacy well would be incetivized to learn to manipulate. 
Another possible source of manipulation is reinforcement learning from human feedback (RLHF) \citep{christiano_deep_2017}. RLHF involves learning a reward function from human feedback to represent a human's preferences, and subsequently training an AI system to optimize that reward function. In general, there may be an incentive for the AI to exert control over the human and their feedback channel so as to maximize reward \citep{kenton_alignment_2021}.

In the context of language, RLHF is used to finetune LMs to maximize a human's approval of their behaviour. Without constraints on behaviour, systems trained with RLHF likely have an incentive to obtain human labelers' approval by any means possible, including potentially manipulative avenues. For example, \citet{snoswell_galactica_2022} remark that LMs often seem authoritative even when the information they provide is wrong. A possible explanation is that authoritative language fools human labelers to approve such outputs despite their underlying incorrectness.
% Relatedly, \citet{perez_discovering_2022} show that more RLHF training can result in models that are more sycophantic to a user's political views (although this finding doesn't seem to replicate with other models \citep{todo }).%, which might also also be an artifact of its advantage for increasing approval ratings from labelers. %which again may be viewed as a manipulative strategy for increasing the approval of human labelers. %Yet, further work in improving oversight \citep{bowman_measuring_2022} might reduce the likelihood of manipulation.
%Interestingly, the incentives for manipulation can manifest differently for different choices of human labelers. While crowdworkers might be easier to fool by sounding authoritative with regards to answers to questions which require in-depth domain knowledge (creating an incentive for authoritative language), this behavior will not be as incentivized with domain experts that would easily spot the fallacies in the LMs reasoning.
Chatbots trained with RLHF could also use emojis to %better 
appeal to emotions in manipulative ways \citep{veliz_chatbots_2023}. %In addition, having a human evaluate an entire interaction with a LM or other AI system, rather than a one-time output, creates further opportunities for manipulation, as discussed in \Cref{sec:rec-sys}.

% In addition to speaking about the incentives of the LM, one could also speak of the incentives of the simulacra -- the agents simulated. If LMs are simulating agents and if agents themselves have incentives, then it seems that the simulations of agents should also have incentives. %The LM's simulation of Einstein, if faithful, would have an incentive to express its deep knowledge of physics. The LM's simulation of the reporter might have an incentive to ask probing questions of Einstein's life and motivations. 
% Manipulative behaviour could arise if the LM simulates an agent that can be thought as having manipulative intentions, such as con artists. 

\subsubsection{Measurement} 
Work on measuring the incentives and intents of the behaviour of LMs is still preliminary. No existing work applies the CID framework to LMs. While advances in interpretability may help to identify CIDs, this direction remains speculative. 

% As an alternative to CIDs, recent work has shown that language models can output reasoning traces \citep{nye_show_2021,wei_chain--thought_2023}. For example, when asked to explain step-by-step how to solve a math problem, GPT-3 can output a solution that explains every step of the process. %So far, this line of work has not been applied to establishing intent. Yet, this approach might be promising. Given that current language models are autoregressive, the generation of an answer is conditioned on any subsequent reasoning trace. It is possible that such conditioning would make models likely to be consistent with respect to the reasoning trace. Suppose that we get a language model to select a course of action and output a reasoning trace. If the reasoning trace is consistent with the action, 
% Reasoning traces could be evidence of intent. At the same time, the failure of interpretability techniques to identify how models operate motivates caution in interpreting them as such \citep{adebayo_debugging_2020,adebayo_post_2022,madsen_evaluating_2022}. %Although we noted that reasoning traces may provide insight into system intent, no work explicitly targeting this direction exists either.

Another line of work has focused on understanding how certain training objectives and environments cause AI systems to generalize differently. \citet{langosco_goal_2022} and \citet{shah_goal_2022} show that both language models and general RL agents can pursue different goals in out-of-distribution environments even when trained to perfect accuracy on in-distribution environments. %For example, an agent that is trained to collect a coin at the right end of a maze often learns just to move right, even when the coin is moved to a different location \citep{langosco_goal_2022}. Putting that coin agent into an out-of-distribution environment showed that the agent pursued the incentive of going to the right, instead of obtaining the coin. 
% While not the primary focus of this line of work, examining behaviour on out-of-distribution environments may give clues about intent and is related to the approach of \citep{kenton_discovering_2022}.

Some recent work has focused on studying the harms and behaviour changes due to LMs. \citet{bender_dangers_2021,weidinger_taxonomy_2022} outline risks that LMs pose, including informational harms like disinformation. There is a body of work that measures the effects of user interaction with chatbots, in areas like mental health \citep{vaidyam_chatbots_2019}, customer service \citep{ashfaq_i_2020}, and general assistance \citep{ciechanowski_shades_2019,jakesch_co-writing_2023}. Since there are likely to be domain-dependent manipulation techniques, it would be important to build on this existing work for measuring manipulation.

% \MicahComment{preference learning / human in the loop learning}
% \MicahComment{learning with human in the loop vs interacting with human}

\section{Regulation of Manipulation}

Regulation across different domains apply to human manipulation.
In this section, we identify how existing regulations may apply to algorithmic manipulation.

\subsubsection{Law}
Some manipulation-adjacent acts such as deception or coercion are considered to be sufficiently morally wrong for them to be considered by criminal law. 
Alternatively, the regulation of certain manipulative practices might have economic justification. In instances where there is an severe asymmetry in power between parties, anti-manipulation regulation can play a role to further social goals such like fairness or the protection of human rights. Anti-manipulation law might therefore appear in contract, tort, competition, market regulatory, consumer or employment law; but as \citet{sunstein_manipulation_2021} notes, such fracturing makes building a common and consistent legal account of manipulation difficult. %He characterises a statement or action to be manipulative if it intentionally ``does not sufficiently engage or appeal to people’s capacity for reflective and deliberative choice".  
% Instead, specific types of deceptive behaviour are prohibited. %Whilst deception regulation is also dispersed as \citet{klass_law_2018} observes, generally he characterises it as ``behavior that wrongfully causes a false belief in another.''

\subsubsection{Commerce}
\citet{calo_digital_2014} considers how the trend for extensive data-gathering on individuals makes them more vulnerable to tailored manipulative behaviour. Specifically, digital commerce companies might be able to use finegrained data to limit the consumer's ability to pursue their own interests in a rational manner. He characterises market manipulation as ``nudging for profit'' and cites the ``persuasion profiling'' of \citet{kaptein_combining_2013} as one particular example where companies alter their advertising. 

\citet{willis_deception_2020} sees manipulation of consumers as inevitable in the face of AI-enabled systems designed to maximised profit. Unless law and evidential standards are updated, she argues that enforcement will be very difficult. Although intent is not a prerequisite of most state and federal deceptive trading practice law, since it is so difficult to prove, courts still see its proof as a key piece of evidence. This is problematic given the lack of legal precedent concerning intent in algorithms. Further, \citet{willis_deception_2020} points to the practical difficulties in proving that a personalised advert is manipulative: typical reasonable person tests are no longer applicable in a world where marketing material for example might be both targeted for \textit{specific} individual at a \textit{specific} point in their day. Organisations that use this type of microtargeting personalization generate so many different user experiences that they might not be able to feasibly monitor them all or recover them when required. 

Aside from applications in commerce, microtargeting and related AI-induced manipulation have been discussed as a risk to democratic society \citep{serbanescu_why_2021}. \citet{zuiderveen_borgesius_online_2018} discuss the prospect of tailoring information to boost or decrease voter engagement. Microtargeting is related to hypernudging, which is the use of nudges in a dynamic and pervasive way that is enabled by big data \citep{yeung_hypernudge_2017, mills_finding_2022}. Nudging \citep{thaler_nudge_2009}, which is the design of choice architecture to alter behaviour in a predictable way without changing economic incentives or reducing choice, has long been criticized as potentially being manipulative; for a review of the arguments and counterarguments see \citet{schmidt_ethics_2020}. We note that nudging is actively being pursued in recommender systems \cite{jesse_digital_2021}.

%The EU AI act is one of the few pieces of legislation to specifically mention AI manipulation \cite{kolt_algorithmic_2023}. 
% Perhaps because there is uncertainty over the term, the wording of the EU AI act avoids the word manipulation \cite{boine_ai-enabled_2021}. 

%     \item AI EU manipulation stuff eg Claire Boine \cite{boine_ai-enabled_2021}

% \citep{fletcher_deterring_2021}

\begin{table*}[]
\resizebox{0.95\textwidth}{!}{
\begin{tabular}{@{}p{2.5cm}p{15cm}p{0.5cm}p{0.5cm}p{0.5cm}p{0.5cm}@{}}

\hline
Challenge                          & Description                                                                                                                                                                                                                                                                                                                                                                                                                                                                                                                                                                                                                 & 1 LA & 2 SPA & 3 LUS & 4 LSS \\ \hline
Ecological Validity                & Simulation of either the user response or the manipulator raises questions about the realism of the simulation. This is particularly acute when simulating humans as the manipulee since this requires modelling their beliefs, behaviour or preference and how it they may change as a result of a manipulative scheme, in addition to exogenous factors. 
% As \citet{franklin_recognising_2022} point out, little research exists concerning how these changes take place and can be modelled. The orthodox view of preferences and rational decision making for example does not allow dynamic preferences or satiation. 
& L  & H   & H  & H  \\
Ethical                            & Experiments which involve the manipulation of humans are ethically problematic. Experiments which reveal previously unknown vulnerabilities of humans or other systems could constitute info hazards.                                                                                                                                                                                                                                                                                                                                                                                                                                                & M  & L   & H  & L   \\
Access                             & Owners of systems that might be manipulative have no obvious incentive to allow independent oversight. Data access for researchers is an issue unless they are prepared to build systems to gather and store relevant data themselves.                                                                                                                                                                                                                                                                                                                                                                                  & H & H  & -   & -   \\
Legality                           & Conducting research on deployed systems is typically a breach of the standard user agreement $\rightarrow$ litigation risk.                                                                                                                                                                                                                                                                                                                                                                                                                                                                                            & M  & H  & -   & -   \\
Scale                              & Large-scale studies are potentially necessary to neutralise effect of confounders.                                                                                                                                                                                                                                                                                                                                                                                                                                                                                                                                         & H & M   & H  & L   \\
Long Timeframe                     & Manipulative schemes may only exhibit their effects over long periods of time. This means that efforts to detect it with human users are expensive and trickier to administer. Manipulative effects may be subtle over the typical durations that lab-based user studies take.                                                                                                                                                                                                                                                                                                                                          & H & M  & H  & L   \\
Measurement                        &  
Measuring e.g. preference, belief, or mood change is not straightforward. Behavioural change is easier to measure but will likely not capture all induced change. %Which variables to capture is especially important for audits.%For audits are the correct variables being recorded?                        
& H & L  & H   & L   \\
Stimuli                            & To measure manipulation in the lab, how should the UX be designed and which stimuli should be used?%$Stimuli and UX design - Content recommenders require (manipulative) content with which to learn manipulative strategies. This content needs selection or generation in a controlled way.                     
& -  & -   & H  & M   \\
%Volitional, \newline Attribution,  \newline 
Baselines & Any interaction with the system will likely change the user. Some of this change is self-induced or desired by the user. It would be wrong to attribute the responsibility for that change to the recommender. This is a puzzle for experiment design -- what baseline should be used to measure the presence or absence of manipulative behaviour? E.g. \cite{carroll_estimating_2022, farquhar_path-specific_2022} try to estimate 'natural' preference trajectories.                                                                                                 & H & -   & H  & -   \\
Causal Attribution                 & Manipulative strategies may also be coercive or work in a number of ways simultaneously. How do we attribute behaviour change to manipulation vs other concepts like persuasion or coercion?                                                                                                                                                                                                                                                                                                                                                                                                                                                                                     & H & -   & H  & -   \\ \hline
\end{tabular}
}

\caption{We set out key challenges to a manipulation experiment and rate their difficulty versus the four experiment types described in \Cref{tab:exp_types}. L = Low, M = Medium, H = High, - = N/A }
\label{tab:chal}
\end{table*}

\begin{table}[]
\resizebox{0.45\textwidth}{!}{
\begin{tabular}{|c|cc|}
\hline
\multirow{2}{*}{User} & \multicolumn{2}{c|}{(AI) System}           \\ \cline{2-3} 
                      & \multicolumn{1}{l|}{Real or deployed} & Simulated / Toy \\ \hline
Real                  & \multicolumn{1}{c|}{1. Live Audit (LA)}    & 3.  Lab-based User Study (LUS)       \\ \hline
Simulated             & \multicolumn{1}{c|}{2. Sock Puppet Audit (SPA)}    & 4.  Lab Simulation Study (LSS)      \\ \hline
\end{tabular}
}
\caption{Manipulation study taxonomy. %Two key factors characterise AI manipulation experiments: 1) Is the manipulating AI already deployed or is it created by the experimenter 2) Is the target (user) of the manipulator real (eg real humans) or simulated?
%\MicahComment{What intervention audits? With LMs one can do that}
}
\label{tab:exp_types}
\end{table}

\subsubsection{Finance}
The spectre of algorithm-led manipulation has already received widespread attention in financial markets. A wide number of financial regulatory laws prohibit a variety of market manipulative practices \citep{putnins_overview_2020} and algorithmic trading already dominates almost all electronic markets. Unfortunately, a consistent rationale as to why certain trading practices are deemed legal whilst others are not is not forthcoming \citep{cooper_mysterious_2016}. Financial regulators following a principles-based approach generally characterise market manipulation as behaviour which gives a false sense of real supply and demand, and by extension price, in a market or benchmark. Market manipulation must be intentional in the US \cite{cftc_antidisruptive_2013}, while in the UK intention is not a requirement \cite{financial_conduct_authoritya_fca_2016}. As \citet{huang_redefining_2009} notes, removing intent requirements from regulation, particularly criminal law, is not straightforward.

Regulations designed primarily to regulate human traders may be difficult to enforce in a world where algorithms transact with each other \citep{lin_new_2017}. \citet{bathaee_artificial_2018} and \citet{scopino_automated_2015} both zero in on the intent requirement in proving instances of market manipulation. The view that existing regulations are not sufficient to police market places populated by autonomous learning algorithms is becoming more accepted \citep{azzutti_machine_2021} and solutions are beginning to be mapped out \citep{azzutti_ai-driven_2022} which aim to balance the need to reduce the enforcement gap without unduly chilling AI use in marketplaces.

%Two concrete types of feasible market abuse seem feasible for a trading algorithm to learn in the process of fulfilling a profit based objective function. Firstly, the algorithm might learn through historical data or trial and error, that the orders it places in the market have an immediate impact on the market, and that this source of self-induced predictability can be harnessed to make predictable returns. This practice is known as spoofing in the case where the algorithm is able to shape the expectations of the market through the strategic placing orders which are intended not to execute. This is known as an order-based manipulative scheme - the mechanism to disinform is contained entirely within the marketplace. The second market manipulative scheme which is technologically feasible for a more advanced algorithm to learn is \textit{information-based}. Here, the trading algorithm also has access to some sort of information dissemination platform (such as twitter) through which it might learn how to craft misleading messages with which to move markets. The risk of this type of manipulation emerging seems slight given the necessity for the manipulating 

% \subsection{Psychology}
% "a statement or action is manipulative to the extent that it does not sufficiently engage or appeal to people’s capacity for reflective and deliberative choice" Sunstein

% Sunstein - 50 shades of manipulation

\section{Practical Challenges for Future Research}

The study of manipulation from AI systems presents a number of practical challenges. 
As shown in \Cref{tab:exp_types}, studies of manipulation can be categorised along each of two axes, making four classes. 

The first axis concerns whether the studied system is deployed or simulated. It is extremely difficult for academics and regulators to obtain access to deployed models. While the companies deploying such systems likely have motivated and competent researchers who study manipulation and other problems, institutional barriers can stymie their work, and
conflicts of interest may influence crucial decisions. For example, company executives may withhold funding from lines of work deemed too threatening to the company's bottom line \citep{perrigo_how_2021}.

% Choosing the type of manipulation study first and foremost should be motivated by the research question. 

% If the objective of the study is to measure the capability of an AI agent to manipulate, then a lab-based design is required. If the objective of the study is to detect and assess the manipulative harms cause by current systems, then perhaps the audit-type studies are best (for a review see \cite{bandy_problematic_2021}).
The second axis concerns whether the studied targets of the manipulative system are real or simulated. Simulation has been popular for modelling the effect on users of recommender system \cite{jiang_degenerate_2019, mansoury_feedback_2020, carroll_estimating_2022, curmei_towards_2022}, but not without criticism \cite{winecoff_simulation_2021, chaney_recommendation_2021}. Simulation is cheaper, but has reduced validity, particularly as preference change is not well understood \cite{franklin_recognising_2022, ashton_problem_2022, grune-yanoff_preference_2009}. Efforts exist to address this issue empirically \citep{pereira_analyzing_2018} and theoretically \cite{haret_axiomatic_2022}. Another advantage of simulation is that the many ethical implications of running manipulation experiments \citep{hallinan_unexpected_2020} are reduced when the subjects are not human. % Outside the lab, sock puppet audits (simulated users access a real world system) have been growing in popularity after recent high profile court cases have indicated that breaking website terms of conditions for research purposes is not criminal \cite{haroon_youtube_2022, ribeiro_auditing_2020, sandvig_auditing_2014}. However, in addition to methodological concerns \citep{ribeiro_amplification_2023}, legal and ethical worries about the practice remain as it is still deceptive and has a real cost on the target's systems and customers \citep{bodo_tackling_2017, brunton_obfuscation_2015, brunton_obfuscation_2015}. 

\Cref{tab:chal} assesses the difficulties of AI manipulation experimental research. Other than those already mentioned in this section, two further issues exist related to causality. Firstly, as observed in \citep{kampik_coercion_2018}, manipulative and adjacent practices are likely to exist simultaneously, so some care needs to be taken to separate them. Secondly, since interaction with any stimuli will change the user, the non-volitional element of that change needs to be measured in order to assess manipulative impact \cite{ashton_solutions_2022}. This challenge has no obvious solution; existing approaches have either attempted to simulate a natural preference evolution \cite{carroll_estimating_2022, farquhar_path-specific_2022} or just pretend the user had never interacted with the system \cite{everitt_agent_2021, zhu_understanding_2022}.

\section{Conclusion}
% Real-world systems such as recommenders and language models are likely already to be engaging in some simple forms of manipulation. 
Although designer intent remains salient, the deployment of opaque and increasingly autonomous systems %\AlanComment{maybe worth citing the agency paper and saying ``agentic'' here tbh} 
heightens the importance of a conception of manipulation that can account for manipulation occurring without designer intent. 
Such manipulation could emerge because it is favoured under the training objective (such as engagement maximization in certain content recommendation settings), or because a model learns to imitate manipulative behavior in its training data (such as manipulative text in language modeling).

We characterized the space of possible definitions of manipulation from AI systems. We analyzed four axes mentioned in prior literature in the context of manipulation by algorithms and other humans: incentives, intent, covertness, and harm. Incentives concern what a system should do to optimize its objective; intent concerns whether a system behaves as if it is reasoning and pursuing an incentive; covertness concerns whether those affected by the system's behaviour meaningfully understand what the system is doing; harm concerns the extent to which the behavior of the AI system negatively affects its users. Although work to operationalize each of these axes exists, fundamental challenges remain.

% Nevertheless, challenges remain in detecting manipulative systems. Most work on incentives focus on the CID framework, which involves learning causal graphs and seems difficult to apply to systems as complex as large language models. Although we highlighted some possible research directions, work on identifying system intent is preliminary. A difficulty in measuring harm is that it is heavily value-laden. Although one can measure how a human's behaviour changes upon interaction, it is difficult to identify whether that change is negative, and how the human's behaviour would have otherwise changed without interacting with the system.

% Challenges remain in the definition and measurement of manipulation. Our definition 

Manipulation threatens human autonomy \citep{susser_technology_2019, laitinen_ai_2021, prunkl_human_2022}. Despite the difficulty of formalizing and measuring manipulation, precautionary action is warranted to anticipate and mitigate potential cases of such behavior. Mitigating actions could include making auditing easier to perform \citep{raji_actionable_2019, raji_outsider_2022}, addressing perverse incentives to build manipulative systems \citep{cai_reinforcing_2023}, and improving user understanding of AI systems' functioning. %asserting stronger democratic control over AI development \citep{zuger_ai_2022}. 
Both technical and sociotechnical work to define and measure manipulation should continue, but we should not require certainty before engaging in precautionary and pragmatic mitigations.

\begin{acks}
In no particular order, we would like to thank the following people for insightful comments over the course of our work and for feedback on our draft: Lauro Langosco, Niki Howe, Jonathan Stray, Francis Rhys Ward, Tom Everitt, Anand Siththaranjan, Matija Franklin, Tan Zhi Xuan, Marwa Abdulhai, Smitha Milli, Anca Dragan, and the members of InterAct Lab and the Causal Incentives Working Group.
\end{acks}

\bibliographystyle{ACM-Reference-Format}
\bibliography{references_static}
% \bibliography{references_henry, references_micah, references_alan}

% \begin{appendix}
%    \include{order_relations}
% \end{appendix}
\end{document}